\documentstyle[12pt]{article}
\newcommand{\beq}{\begin{equation}}
\newcommand{\eeq}{\end{equation}}
\newcommand{\beqn}{\begin{eqnarray}}
\newcommand{\eeqn}{\end{eqnarray}}
\newcommand{\1}{\Lambda}
\newcommand{\2}{\Lambda_E}
\def\vdir{v\kern-7.8pt\Big{/}}
\def\pdir{p\kern-7.8pt\Big{/}}

\begin{document}

\title{A new formulation of the effective theory for heavy particles}
\author{$~~~$}
\date{}
\maketitle
\centerline{ \it U.~Aglietti$~^*$ }

\vskip 0.7truecm

\noindent
INFN, Sezione di Roma, P.le
Aldo Moro 2, 00185 Roma, Italy
\vskip 0.7truecm
\centerline{\it S. Capitani}
\vskip 0.7truecm

\noindent
Dipartimento di Fisica, Universit\`a di Roma ``La Sapienza'',
P.le Aldo Moro 2, 00185 Roma, Italy

\bigskip
\begin{abstract}

\noindent
We derive the effective theories for heavy particles with a functional integral
approach by integrating away the states with high velocity and with high
virtuality. This formulation is non-perturbative
and has a close connection with the Wilson
renormalization group transformation.
The fixed point hamiltonian of our transformation coincides with
the static hamiltonian and
irrelevant operators can be identified with the usual
$1/M$ corrections to the static theory.
No matching condition has to be imposed between
the full and the static theory operators with our approach.
The values of the matching
constants come out as a dynamical effect of the renormalization
group flow.

\end{abstract}
\bigskip
\bigskip

\noindent
email address:

\noindent
aglietti@theory.caltech.edu,

\noindent
capitani@vaxrom.roma1.infn.it

\bigskip

\noindent
$~^*$ at present at Caltech, Pasadena, CA 91125, USA.
\newpage

\section{Introduction}

Effective theories for heavy particles are a powerful tool to study
heavy flavor physics. In recent years, many properties of the
spectrum and of the decays of heavy
hadrons have been understood \cite{neub}.
The idea underlying the effective theory approach
is that typical (i.e. after renormalization) momentum transfers
between quarks and gluons in a
hadron are of the order of the $QCD$ scale
$\1_{QCD} \sim 300~MeV$. As a consequence,
relativistic processes for a ``heavy
quark'' $Q$, with a mass $M\gg\Lambda_{QCD}$, are suppressed.
The heavy quark is almost at rest and on-shell,
and chromomagnetic interactions are weak \cite{eita}.
Furthermore, the creation
of heavy $Q\overline{Q}$ pairs may be neglected.

In general,
we consider processes where {\it heavy} particles are
subjected to {\it soft} interactions.
By `soft' we mean that the energy transfer $\epsilon$ and the
momentum transfer $\vec{q}$ to the heavy particle $H$ are much less
than its mass $M$:
\beq
\mid\epsilon\mid, ~\mid\vec{q}\mid~\ll~ M .
\eeq
Before the scattering $H$ is on-shell and, let us say, at
rest. After soft interactions, $H$ will remain
essentially on-shell and will
acquire a very small velocity.
The relevant states for the dynamics of $H$
will be those with 4-momentum
around the rest momentum
\beq
r~=~(M,\vec{0}) .
\label{eq:expa}
\eeq
The effective theories  are derived generally by
means of an expansion in $k/M$ of the lagrangian \cite{fw},
where $k$ is the residual momentum defined as
\beq\label{eq:shift}
k~=~p-r ,
\eeq
and $p$ is the momentum of the heavy particle.
This expansion is of classical nature and the effects of
quantum fluctuations are taken into account imposing a set of
matching conditions to the effective theory
operators \cite{eichil}.

We reformulate the effective theories for heavy particles with a
functional integral approach \cite{mannel}.
We start with a field theory
with a cut-off $\Lambda$ much larger than the heavy particle
mass,
\beq\label{eq:continuo}
\Lambda~\gg~M ,
\eeq
and we integrate away:
\begin{enumerate}
\item[$i)$] the states with a large velocity, for which the
spatial momentum $\mid\vec{p}\mid\sim M$
or greater than that;
\item[$ii)$] the states which are highly
virtual, i.e. the states with an invariant mass
$p^2\ll M^2$ or $p^2\gg M^2$.
\end{enumerate}
We leave in the effective action $S_{eff}$ the states with momenta
$p$ inside a small
neighborhood of the on-shell momentum of
eq.~(\ref{eq:expa}) (see fig.1):
\beq
(p_0-M)^2~+~\vec{p}{~^2}~~\leq~~\Lambda_E^2 ,
\eeq
where $\Lambda_E$ is the ultraviolet cut-off of the effective theory,
well below the heavy particle mass:
\beq
\Lambda_E~\ll~M .
\eeq
In this way we confine ourselves to the low-energy domain.
The effective action  $iS_{eff}$ is defined as:
\beqn\label{eq:minep}
&&\exp\{~iS_{eff}[\phi(k); 0<k^2<\Lambda_E^2]~\}\doteq
\\ \nonumber
&&~~~~~~~~~~~\int\prod_{\Lambda_E^2<k^2<\Lambda^2}d\phi(k)
{}~\exp\{~iS[\phi(l); 0<l^2<\Lambda^2]~\} .
\eeqn
There exists a different definition of $S_{eff}$
with respect to that one
given in eq.~(\ref{eq:minep}).
We may also consider the construction
of the effective action for heavy particles as a scale transformation.
The system is viewed at progressively larger scales with respect to the
Compton wavelength of the heavy particle, $\lambda_C / 2\pi=1/M$.
We include therefore in the definition
of the effective action, in addition
to the integration of the high-energy modes, also a rescaling
of the residual momenta and of the fields.
This second formulation has a closer relation with the
Wilson renormalization group ($RG$)
transformation \cite{Wilson} and consists of the following steps:
\begin{enumerate}
\item[1)] We lower the ultraviolet cut-off
of the effective action by a
factor $s$, i.e. from $\Lambda_E$ to $\Lambda_E/s$. The modes
with momenta between $\Lambda_E$ and $\Lambda_E/s$ are integrated:
\beqn\label{eq:step1}
&&\exp\{~iS_{eff}[\phi(k); 0<k^2<(\Lambda_E/s)^2]~\}\doteq
\\ \nonumber
&&~~~~~~~~~~~\int\prod_{(\Lambda_E/s)^2<k^2<\Lambda_E^2}d\phi(k)
{}~\exp\{~iS[\phi(l); 0<l^2<\Lambda_E^2]~\} .
\eeqn
\item[2)] The (residual) momentum is rescaled according to:
\beq\label{eq:step2}
k'~=~sk .
\eeq
Notice that one does not rescale the energy of the heavy quark, but
the energy minus the mass.
\item[3)] The heavy particle field $\phi$ is rescaled in such a way that
the kinetic operator takes a unit coefficient:
\beq\label{eq:step3}
\phi'(k')~=~\zeta~ \phi(k) .
\eeq
\end{enumerate}
This transformation can be iterated many times.
At each step we lower the cut-off $\Lambda_E$ by a factor $s$,
i.e. from $\Lambda_E$ to $\Lambda_E/s$,
  from $\Lambda_E/s$ to $\Lambda_E/s^2$,
etc., and we generate a sequence
of effective actions
$S_{eff},~S_{eff}^{(1)},~S_{eff}^{(2)},\ldots~$.

\noindent
Expanding the effective action $S_{eff}$ in a basis of local operators
$O_n$,
\beq
S_{eff}~=~\sum_n c_n O_n ,
\eeq
where $c_n$ are given coefficients,
the $RG$ transformation has the following representation:
\beq
c_n^{(l+1)}~=~f_n(c_1^{(l)},c_2^{(l)},...,c_k^{(l)},...;s)
{}~~~~n=1,2,3,...,k,...~~.
\label{eq:exflow}
\eeq
These equations are transformed into differential equations by
setting $s=1+\delta s$ with $\delta s\ll 1$, and expanding up to
first order in $\delta s$.

\noindent
The fixed point action $S^*$ (if any) describes particles which can change
their momenta with respect to the rest
momentum in eq.~(\ref{eq:expa}) by a tiny
amount only, i.e. it describes particles which are essentially static.
$S^*$ therefore must be identified with the static action.
A fixed point is a solution of:
\beq
c_n^*~=~f_n(c_1^*,c_2^*,...,c_k^*,...;s)~~~~~~~~~n=1,2,3,...,k,...~~.
\eeq
Finding a fixed point for our $RG$ transformation can be
considered as a rigorous proof of the existence of the static limit
in quantum field theory.

Let us make some remarks about the differences between the
formulation specified
by eqs.~(\ref{eq:step1}-\ref{eq:step3}) and the usual one.

With our formulation we do not deal with lagrangians
but directly with the functional integral:
quantum fluctuations are taken into account from the very beginning.
In the usual approach, one performs an expansion in $1/M$ of
the (relativistic) lagrangian ${\cal L}$ for a heavy particle to derive
the static lagrangian ${\cal L}_{st}$.
Tree level amplitudes computed with ${\cal L}$ and with ${\cal L}_{st}$
coincide at order $(1/M)^0$, implying no need of lowest order matching.
Loop diagrams are sensitive to high energy
physics and turn out to have a different value
in the full and in the
static theory. This implies that
quantum fluctuations spoil tree level matching, and one is forced to
redefine the matching constants in order to include their effects.
With our approach, there is not any matching condition to
impose: one has simply to integrate away the off-shell modes
and look at the modifications of the coefficients of the operators.

Unlike classical physics or quantum mechanics,
the static approximation is singular in quantum field theory.
The reason is that the heavy particle mass $M$ acts as a (physical)
ultraviolet cut-off in many processes. Logs of $M$, for example,
appear in one loop corrections to heavy currents \cite{eichil}.
In the static theory, the heavy particle mass is removed and
the resulting divergence is regulated by $\Lambda_E$.
The cut-off of the effective theory, therefore, does not regulate
only the divergences related to mass, field and coupling constant
renormalization, but also those ones induced by the limit
$M\rightarrow\infty$.
With our approach, the singular behaviour of the static approximation
is revealed by an anomalous scaling as we lower progressively
$\Lambda_E$.

Ordinary $RG$ transformation is a tool
to describe {\it soft} interactions
of {\it light} particles in the framework of a field theory.
In this case
the states with large {\it full} energy and with large 3-momenta
are integrated away and the
effective action $S_{eff}$ describes the dynamics
of light particles
with momenta inside a small neighborhood of the
null momentum (see fig.2)
\beq
n~=~(0,\vec{0}) .
\eeq
We see therefore that our
construction of the effective action for heavy
particles specified by eqs.~(\ref{eq:step1}-\ref{eq:step3})
closely parallels Wilson's $RG$ transformation.
The difference is that we integrate away
the states with momenta far from
the rest momentum of a heavy particle instead of the null momentum
(see ref.\cite{Grinst} where an intuitive discussion on the above
point is presented).

\noindent
This paper is organized as follows.

In Sect.~\ref{freeca} we compute the $RG$ transformation
(\ref{eq:step1}-\ref{eq:step3}) in the free case. This is
the simplest case (exactly solvable)
and will serve as an illustration of the
basic ideas and of the formalism.

In Sect.~\ref{simpmod} we construct the $RG$ transformation
for a simple (interacting) model, and we discuss the general
approximations involved.
A perturbative evaluation of the $RG$ transformation
and of its fixed points is also presented.

In Sect.~\ref{match} we compute the effective hamiltonian as
given by eq.~(\ref{eq:minep}) - i.e. without rescaling -
for the  same model,
and we discuss the relation between the original
and the effective hamiltonian.
A perturbative evaluation of the
matching constants is presented and
the relation with ordinary matching theory is treated.

Sect.~\ref{concl} contains the conclusions of our analysis.

\section{Free case}
\label{freeca}

\indent
Let us consider the construction of the effective
action in a very simple case, a free scalar of mass $M$ in
Minkowski space. The case of fermions will be discussed in a
subsequent work.
The action is given by:
\beq
iS~=~i\int_{0}^{\Lambda} \frac{d^Dp}{(2\pi)^D}
     ~\Phi^{\dagger}(p)~[p^2-M^2+i\epsilon]~\Phi(p) .
\eeq
Since what is small is not the energy of the heavy particle,
but the energy minus the mass, it is convenient to express the
action in terms of the residual momentum $k$.
One has:
\beq
iS~=~i\int_0^{\Lambda} \frac{d^Dk}{(2\pi)^D}
     ~\phi^{\dagger}(k)~[k_0+\frac{k^2}{2M}+i\epsilon]~\phi(k) ,
\label{eq:mine}\eeq
where we have defined an `effective field' $\phi$ such that:
\beq
\phi(k)~\doteq~\sqrt{2M}~\Phi(r+k) .
\eeq
The difference between the domains of integration of $p$ and $k$
has been neglected because of
the condition stated in eq.~(\ref{eq:continuo}).
In the free case, the functional integration is trivial because there
is no coupling between high-momentum and
low-momentum modes, and gives an
effective action of the same form as the original one,
with a smaller cut-off $\Lambda_E$:
\beq
iS_{eff}~=~i\int_{0}^{\Lambda_E}
\frac{{\rm d}^Dk}{(2\pi)^D}~
\phi^{\dagger}(k)~
[k_0+k^2/2M+i\epsilon]~\phi(k) .
\label{eq:compl}\eeq

\noindent
In order to understand the relevance of
the operators entering in $S_{eff}$,
let us compute the $RG$ transformation as defined by
eqs.~(\ref{eq:step1}-\ref{eq:step3}).
\begin{enumerate}
\item[1)] Integrating the modes with  momenta between
$\Lambda_E$ and $\Lambda_E/s$, the effective action becomes:
\beq
iS_{eff}'~=~i\int_{0}^{\Lambda_E/s}
\frac{{\rm d}^Dk}{(2\pi)^D}~
\phi^{\dagger}(k)~
[k_0+k^2/2M+i\epsilon]~\phi(k) .
\eeq
plus a constant which does not affect the value of the correlation
functions.
\item[2)] Rescaling the momenta according to eq.~(\ref{eq:step2}),
the effective action reads:
\beq
iS_{eff}'~=~is^{-(D+1)}~\int_{0}^{\Lambda_E}
\frac{{\rm d}^D k'}{(2\pi)^D}~
\phi^{\dagger}(k'/s)~
[k_0'+\frac{1}{s}\frac{k'^2}{2M}+i\epsilon]~\phi(k'/s) .
\eeq
\item[3)] Rescaling the field $\phi$ in such a way that the kinetic
operator takes a unit coefficient,
\beq
\phi'(k')=s^{-(D+1)/2}\phi(k'/s) ,
\eeq
the effective action looks finally:
\beq
iS_{eff}'~=~i\int_{0}^{\Lambda_E}\frac{{\rm d}^D k}{(2\pi)^D}
\phi'^{\dagger}(k')~[k_0'+\frac{1}{s}
\frac{k'^2}{2M}+i\epsilon]~\phi'(k') .
\eeq
\end{enumerate}
We iterate now the transformation many times, say $n$ times; each
iteration the term with $k^2$ is multiplied by a factor $1/s$
and its coefficient reduces exponentially with $n$.
There exists therefore a fixed point action $S^*$, which is given by:
\beq
iS^*~=~i\int_{0}^{\Lambda_E}\frac{{\rm d}^D k}{(2\pi)^D}
\phi^{\dagger}(k)~[k_0+i\epsilon]~\phi(k) .
\eeq
This is the familiar static action. Note that the mass $M$ of the
heavy particle disappears in the fixed point action $S^*$.
In other words, the actions of heavy particles with different masses
lie in the same universality class with respect to the $RG$
transformation specified by eqs.~(\ref{eq:step1}-\ref{eq:step3}).
This is the well-known flavor symmetry of the
static theory \cite{geo}.
The term $k^2/2M$ is an irrelevant operator with
respect  to our $RG$ transformation.
For high level computations
this operator may be introduced in the dynamics as an insertion,
by means of an expansion of the form:
\beq\label{eq:sde}
e^{ iS_{eff} }~=~e^{ iS^* }~
\Big[1+i\int_0^{\Lambda_E}\frac{ {\rm d}^Dk }{ (2\pi)^D }
\phi^{\dagger}(k)\frac{k^2}{2M}\phi(k)+O(1/M^2)\Big] .
\eeq
According to the short-distance expansion (\ref{eq:sde}),
the heavy scalar propagator is given by:
\beqn\label{eq:propexp}
i\Delta(k)&=& \frac{i}{k_0+i\epsilon}+\frac{i}{k_0+i\epsilon}
\Bigg(i\frac{k_0^2}{2M}-i\frac{\vec{k}^2}{2M}\Bigg)
\frac{i}{k_0+i\epsilon}+O(1/M^2)
\\ \nonumber
&=&\frac{i}{k_0+i\epsilon}\Bigg(1-\frac{k_0}{2M}\Bigg)
+\frac{i}{k_0+i\epsilon}
\frac{-i\vec{k}^2}{2M}\frac{i}{k_0+i\epsilon}+O(1/M^2) .
\eeqn
The round bracket contains a contact term,
which is the effect of order
$1/M$ of the antiparticle pole, i.e. of the states
with a very high virtuality. This term
can be reabsorbed with a redefinition
of the operators containing the heavy field $\phi$:
\beq
\phi~\rightarrow~\Bigg(1-i\frac{\partial_0}{4M}\Bigg)\phi .
\eeq
The second term of the last member in eq.~(\ref{eq:propexp}) is
the familiar kinetic energy operator.
We conclude therefore
that the irrelevant operators with respect
to our $RG$ transformation
coincide with the operators of order $1/M$
in the expansion of the lagrangian.

The $RG$ transformation constructed above
can be easily extended to describe infrared
interactions of heavy particles with a fixed velocity $v^{\mu}$
\cite{geo,iw}.
The only difference
is that in the latter case we must leave in the effective action
the states with momenta in a small neighborhood of the momentum $Mv$
instead of the rest momentum of eq.~(\ref{eq:expa}).

\section{A simple model}
\label{simpmod}

We present in this section the construction of the effective hamiltonian
in an interacting case.
We consider a simple model, in euclidean space,
whose dynamics is determined by
the hamiltonian:
\beq
{\cal H}~=~\partial_{\mu}\Phi^{\dagger}\partial_{\mu}\Phi
+M^2\Phi^{\dagger}\Phi+\frac{1}{2}\partial_{\mu}a\partial_{\mu}a
+\frac{1}{2}m^2a ^2 + g\Phi ^{\dagger}\Phi a ,
\eeq
where $\Phi$ is a heavy scalar field with mass $M$ and $a$ is a light
scalar field with mass $m$.
\noindent
This theory is superrenormalizable in four dimensions $D=4$, because the
coupling $g$ has the dimension of a mass, but this does not matter
for the following considerations. We do not deal with the continuum
limit of the model (i.e. with the limit $\Lambda\rightarrow\infty$)
but rather with its infrared properties.
This is indeed a model for $QED$ at
small energies compared to the electron mass.

We are interested in the soft interactions
between these particles; we introduce therefore an effective theory
where we integrate away the
states with high-energy for the light particle
and the states far from the mass-shell state of eq.~(\ref{eq:expa})
for the heavy particle.
We select a cut-off $\Lambda_E$ so that:
\beq\label{eq:gerar}
m^2\ll \Lambda_E^2\ll M^2 .
\eeq

\noindent
Let us compute the $RG$ transformation as specified by
eqs.~(\ref{eq:step1}-\ref{eq:step3}).

1) After making a shift like in eq.~(\ref{eq:mine}), the integration
of the high energy modes is realized according to:
\beqn\nonumber
& &~~~~~~~\exp\Big\{-H_{eff}'
[~\phi(k),a(l);0<k^2,l^2<(\Lambda_E/s)^2]~\Big\}=
\\ \nonumber
& &\int\prod_{(\frac{\Lambda_E}{s})^2<k^2,l^2<\Lambda_E^2}
{\rm d}\phi(k){\rm d} a(l)
\exp\Big\{-\int_0^{\Lambda_E} \frac{d^D k}{(2\pi)^D}
\phi^{\dagger}(k)\Big[ik_4+\delta M+\frac{k^2}{2M}\Big]\phi(k)
\\ \nonumber
&-&\int_0^{\Lambda_E}\frac{d^Dl}{(2\pi)^D}~
a(-l)~[l^2+m^2]~a(l)+
\\
&-&\lambda\int_0^{\Lambda_E}
\frac{d^Dk_1}{(2\pi)^D}\frac{d^Dk_2}{(2\pi)^D}
\frac{d^Dk_3}{(2\pi)^D}(2\pi)^D\delta(k_1-k_2-k_3)
\phi^{\dagger}(k_1)\phi(k_2)a(k_3)\Big\} ,
\label{eq:long}\eeqn
where $\lambda=g/2M$ is a dimensionless coupling constant in $D=4$,
and a bare mass
term $\delta M$ has been introduced in the heavy particle lagrangian.

\noindent
The integrated modes induce non-local couplings
in the effective hamiltonian,
as well as interactions between an arbitrary
number of particles.
$H_{eff}'$ therefore is non local
and non polynomial even though the original hamiltonian is
renormalizable:
\beqn \nonumber
& &H_{eff}'~=~\int_0^{\Lambda_E/s} \frac{d^D k}{(2\pi)^D}~
\phi^{\dagger}(k)~
\Big[ik_4+\delta M+\frac{k^2}{2M}-\Sigma_H(k)\Big]~\phi(k)
\\ \nonumber
&+&\int_0^{\Lambda_E/s}
\frac{d^Dl}{(2\pi)^D}~ a(-l)~\Big[l^2+m^2-\Sigma_L(l)\Big]~a(l)+
\\ \nonumber
&+&\int_0^{\Lambda_E/s}
\frac{d^Dk_1}{(2\pi)^D}...
\frac{d^Dk_3}{(2\pi)^D}(2\pi)^D\delta(k_1...-k_3)
\Big[\lambda-V_{\phi\phi }^a(k_1,k_2,k_3)\Big]
\phi^{\dagger}(k_1)\phi(k_2)a(k_3)
\\ \nonumber
&+&\int_0^{\Lambda_E/s}
\frac{d^Dk_1}{(2\pi)^D}...
\frac{d^Dk_4}{(2\pi)^D}
(2\pi)^D\delta(k_1...-k_4)
V_{\phi\phi }^{aa}(k_1,...,k_4)
\phi^{\dagger}(k_1)\phi(k_2)a(k_3)a(k_4)
\\
&+&\ldots~~,
\label{eq:long2}\eeqn
where we have separated the tree level coupling from the loop
correction in the $\phi\phi a$ vertex,
i.e. $V_{\phi\phi}^a$ starts
at order $\lambda^3$.

An exact evaluation of $H_{eff}'$ is clearly an hopeless task
and some approximation is needed.
Assuming a basis of local operators, this involves a truncation
of the operator series generated
by eq.~(\ref{eq:long}). The justification
of this approximation relies on scale considerations
and will be discussed in 2).

2) Rescaling the full momenta of
the light particle and the residual
momenta of the heavy particle according to eq.~(\ref{eq:step2}),
brings the effective hamiltonian into the form:
\beqn \nonumber
& &H_{eff}''=\int_0^{\Lambda_E} \frac{d^D k'}{(2\pi)^D}~
s^{-(D+1)}\phi^{\dagger}(\frac{k'}{s})~
\Big[ik'_4+s\delta M+\frac{1}{s}\frac{{k'}^2}{2M}
-s\Sigma_H(\frac{k'}{s})\Big]~\phi(\frac{k'}{s})
\\ \nonumber
&+&\int_0^{\Lambda_E}
\frac{d^Dl'}{(2\pi)^D}s^{-(D+2)}~
a(-l'/s)~\Big[{l'}^2+s^2m^2-s^2\Sigma_L(l'/s)\Big]~a(l'/s)+
\\ \nonumber
&+&\int_0^{\Lambda_E}
\frac{d^D{k'}_1}{(2\pi)^D}...
\frac{d^D{k'}_3}{(2\pi)^D}(2\pi)^D\delta(k'_1...-k'_3)
s^{-2D}\Big[ \lambda - V_{\phi\phi}^a
(\frac{k'_1}{s},\frac{k'_2}{s},\frac{k'_3}{s})\Big]
\phi^{\dagger}(\frac{k'_1}{s}) \cdots a(\frac{k'_3}{s})
\\ \nonumber
&+&\int_0^{\Lambda_E}
\frac{d^Dk'_1}{(2\pi)^D}...\frac{d^Dk'_4}{(2\pi)^D}
(2\pi)^D\delta(k'_1...-k'_4)s^{-3D}
V_{\phi\phi }^{aa}(\frac{k'_1}{s},...,\frac{k'_4}{s})
\phi^{\dagger}(\frac{k'_1}{s}) \cdots a(\frac{k'_4}{s})
\\
&+&\ldots~~.
\label{eq:longp}\eeqn
The original limit of integration,
$\Lambda_E$, has been restored.

\noindent
Let us discuss now the main
 approximation in the evaluation of $H_{eff}$.
We consider first the free field case ($\lambda=0$).
As it stems from eq.~(\ref{eq:longp}),
the mass dimensions of the fields $\phi(k)$  and $a(l)$
are given respectively by:
\beq
-\frac{D+1}{2},~~~~-\frac{D+2}{2} .
\eeq
The dimension of a derivative operator is one.
Then, the dimension of a composite operator
$O$ (integrated over space)
containing $n_{\phi}$ $\phi$-fields , $n_{a}$ $a$-fields and
$n_{\partial}$  derivatives is given by:
\beq
d_O~=~-D+n_{\phi} \frac{D-1}{2} + n_a\frac{D-2}{2} + n_{\partial} .
\eeq
Each $RG$ iteration, the operator $O$ is multiplied by the factor
$s^{-d_O}$.  Therefore, an operator with a positive mass dimension,
$d_O>0$, has a weight which reduces exponentially with the number
of iterations, and can be neglected at leading level.
In the case with interactions ($\lambda\neq 0$), we assume that their
effects are small, in the sense
that anomalous scaling does not overwhelm the canonical one.
Also in the latter case, therefore, we can neglect the
operators with a positive mass dimension.

The only relevant operators
(i.e. with $d_O\leq 0$) are those already present
in the original hamiltonian,
\beq
\phi^{\dagger}\phi,~~~\phi^{\dagger}k_4\phi,~~~
a^2,~~~al^2a,~~~\phi^{\dagger}\phi a ,
\eeq
plus
\beq
\phi^{\dagger}(k)k_i\phi(k)~~~~~~~~~~~~~{\rm for}~~i=1,2,3 .
\eeq
The latter are indeed the operators which enter in the Georgi lagrangian
\cite{geo}.
We will see in section (\ref{pertev})
that they are not induced in $H_{eff}$
for the static case in a regularization preserving parity.

At the leading level, therefore,
the $RG$ transformation amounts to a renormalization of the
original couplings only:
\beqn \nonumber
&&H_{eff}''\cong\int_0^{\Lambda_E} \frac{d^D k'}{(2\pi)^D}
s^{-(D+1)}\phi^{\dagger}(\frac{k'}{s})
\Bigg[
\Bigg(1+i
\left(\frac{\partial\Sigma_H}{\partial k_4}\right)_0\Bigg)ik'_4+
s\Big(\delta M-\Sigma_H(0)\Big)\Bigg]\phi(\frac{k'}{s})
\\ \nonumber
&+&\int_0^{\Lambda_E}
\frac{d^Dl'}{(2\pi)^D}s^{-(D+2)}
a(-\frac{l'}{s})\Bigg[
\Bigg(1-\left(\frac{\partial\Sigma_L}{\partial l^2}
\right)_0\Bigg)
{l'}^2+s^2\Big(m^2-\Sigma_L(0)\Big)\Bigg]a(\frac{l'}{s})+
\\ \nonumber
&+&\int_0^{\Lambda_E}
\frac{d^Dk'_1}{(2\pi)^D}...
\frac{d^Dk'_3}{(2\pi)^D}(2\pi)^D\delta(k'_1...-k'_3)
s^{-2D}\Bigg[\lambda-V_{\phi\phi }^a(0,0,0)\Bigg]
\phi^{\dagger}(\frac{k'_1}{s})
\phi(\frac{k'_2}{s})a(\frac{k'_3}{s}) .
\\
& &~~~~~
\label{eq:longs}\eeqn

3) We rescale now the heavy and the light particle fields in such
a way that the kinetic operators
$\phi^{\dagger}(k)ik_4\phi(k)$ and
$a(-l)l^2a(l)$ respectively, take unit coefficient:
\beqn  \nonumber
\phi'(k')&=&s^{-(D+1)/2}
\Bigg[1+i
\left(\frac{\partial\Sigma_H}{\partial k_4}\right)_0\Bigg]^{1/2}
            \phi(k'/s)
\\
a'(l')&=&s^{-(D+2)/2}
\Bigg[1-
\left(\frac{\partial\Sigma_L}{\partial l^2}\right)_0\Bigg]^{1/2}
a(l'/s) .
\label{eq:zetas}\eeqn
The effective hamiltonian takes finally the form:
\beqn\nonumber
& &H_{eff}'''=\int_0^{\Lambda_E} \frac{d^D k'}{(2\pi)^D}~
{\phi'}^{\dagger}(k')~\Big[ik'_4+ \delta M'\Big]~\phi'(k')+
\\
&+&\int_0^{\Lambda_E}\frac{d^Dl'}{(2\pi)^D}~ a'(-l')~
\Big[l'^2+{m^2}^{'} \Big]~a'(l')+
\\ \nonumber
&+&\lambda'\int_0^{\Lambda_E}
\frac{d^Dk'_1}{(2\pi)^D}\frac{d^Dk'_2}{(2\pi)^D}
\frac{d^Dk'_3}{(2\pi)^D}(2\pi)^D\delta(k'_1-k'_2-k'_3)
{\phi'}^{\dagger}(k'_1)\phi'(k'_2)a'(k'_3) ,
\eeqn
where the new coupling $\lambda'$ and the new mass parameters
$\delta M'$ and ${m^2}^{'}$ are given by:
\beqn\nonumber
\lambda'&=&s^{2-D/2}
\Bigg[1-\frac{V_{\phi\phi}^a(0,0,0)}{\lambda}\Bigg]
\Bigg[ 1+i\left(\frac{\partial\Sigma_H}{\partial k_4}\right)_0
\Bigg]^{-1}
\Bigg[1-
\left(\frac{\partial\Sigma_L}{\partial l^2}\right)_0\Bigg]^{-1/2}
\lambda
\\ \nonumber
\delta M'&=& s
\Bigg[1+i\left(\frac{\partial\Sigma_H}{\partial k_4}\right)_0
\Bigg]^{-1}
\Bigg[\delta M-\Sigma_H(0)\Bigg]
\\
{m^2}^{'}&=& s^2
\Bigg[1-\left(\frac{\partial\Sigma_L}{\partial l^2}\right)_0
\Bigg]^{-1}
\Bigg[m^2-\Sigma_L(0)\Bigg] .
\label{eq:flow}\eeqn

Eqs.~(\ref{eq:flow}) are an approximation of the exact
equations (\ref{eq:exflow}) for the $RG$ flow and,
together with (\ref{eq:zetas}), represent the leading effects
of the $RG$ transformation.
To proceed, we must compute explicitly the functions entering in
eqs.~(\ref{eq:flow}). This is done in the next section with
perturbative methods.
We will search the fixed points
of the truncated eqs.~(\ref{eq:flow})
in Sect.~\ref{fixpo}.

\subsection{Perturbative computation}
\label{pertev}

The construction of the effective hamiltonian
presented in Sect.~\ref{simpmod}
is non-perturbative; it is basically a prescription
for the degrees of freedom
to integrate away.
As such, it does not rely on any Feynman diagram expansion, just
like Wilson transformation.
In deriving it, we assumed only that the effects of
interactions are not so strong
to destroy the distinction between relevant and irrelevant operators
given by the free theory.
This assumption is indeed quite reasonable from
the physical viewpoint, and it can only be checked case by case
with an exact computation.
However, in order to develop an intuitive understanding of the
properties of our transformation, and to make contact with the
usual matching theory, we present in this section a perturbative
evaluation of the effective hamiltonian.

Let us sketch the derivation of the Feynman rules for the evaluation
of $H_{eff}'$ \cite{ma}.
We decompose the fields $\phi$ and $a$ into their low energy and high
energy parts:
\beq\label{eq:decomp}
\phi(k)~=~\phi_L(k)+\phi_H(k),~~~~~~~a(l)~=~a_L(l)+a_H(l) ,
\eeq
where the fields with the subscripts `$L$' or `$H$' contain modes with
momenta less than or above $\Lambda_E/s$ respectively:
\beqn\nonumber
\phi_L(k)&=&\phi(k)~\theta(\Lambda_E/s-k),~~~~~~
                       \phi_H(k)~=~\phi(k)~\theta(k-\Lambda_E/s)
\\ a_L(l)&=&
a(l)~\theta(\Lambda_E/s-l),~~~~~~~~a_H(l)~=~
a(l)~\theta(l-\Lambda_E/s) .
\eeqn
We insert then the decomposition
(\ref{eq:decomp}) in the right hand side of
eq.~(\ref{eq:long}). In the bilinear (free) part of the hamiltonian
there are no couplings between `$L$' and `$H$' fields. On the contrary,
the trilinear (interaction) term generates any possible
couplings between them.
We expand
the right hand side of eq.~(\ref{eq:long}) in powers of $\lambda$
and we compute the functional
integral, which is only over $\phi_H$ and $a_H$.
Each power of $\phi_L$ and $a_L$ is associated to an external leg
of the diagrams, while each bilinear in $\phi_H$ or $a_H$
is associated to an internal line.
The external lines of the diagrams have momenta between zero
and $\Lambda_E/s$ while the internal lines have momenta in the range
$\Lambda_E/s-\Lambda_E$. Loops are integrated in a region where
all the internal lines have momenta
between $\Lambda_E/s$ and $\Lambda_E$.
\noindent
The Feynman rules are:
\beqn
\frac{-i}{k_4-ik^2/2M}&:&{\rm heavy~ scalar~propagator}
\\
\frac{1}{l^2+m^2}&:&{\rm light~ scalar~propagator}
\\
-\lambda&:&{\rm vertex} .
\eeqn
Note the asymmetry between
the heavy and the light propagator related to
the shift in the energy for the massive one; the poles for the
heavy particle and the heavy antiparticle are respectively at:
\beq
k_4~=~i(M\mp\sqrt{\vec{k}^2+M^2}) .
\eeq
By taking the logarithm on both sides
of eq.~(\ref{eq:long}), one sees that
only connected diagrams contribute to $H_{eff}'$.
Furthermore, one particle reducible diagrams do not contribute
to the effective hamiltonian in the low momentum region.

\noindent
The exact determination of $H_{eff}'$
therefore requires to compute and to sum
all the connected diagrams with an arbitrary
number of external legs and with
generic momenta below $\Lambda_E/s$.
Loops  induce in the effective hamiltonian
interactions with an arbitrary number of fields and with an
arbitrary number of derivatives.
As explained in the previous section,
we neglect higher dimension
operators since they are irrelevant,
and we limit to one loop accuracy.

\noindent
Let us go then to the explicit computation of the diagrams.
  From now on we stick to the case $D=4$. At one-loop level,
the one-particle irreducible diagrams are classified by the number
of external heavy and light lines, ($n_H,~n_L$).

\noindent
The self-energy of the heavy scalar ($n_H=2,~n_L=0$) is given by:
\beq
\Sigma_H(k)=-i\lambda^2\int_{D2}\frac{d^4l}{(2\pi)^4}
\frac{1}{l_4-il^2/2M}~\frac{1}{(k-l)^2+m^2} ,
\label{eq:smine}\eeq
where $D2$ is the region of the $l$-space, where both propagators
have momenta between $\Lambda_E/s$ and $\Lambda_E$:
\beq
(\Lambda_E/s)^2~<~l^2,~(k-l)^2~<\Lambda_E^2 .
\eeq
Since ${m^2}^*=0$ (see later),
we can neglect $m^2$ in eq.~(\ref{eq:smine}).

\noindent
The self-energy at zero external momentum is given by:
\beq
\Sigma_H(0)~=~\frac{\lambda^2}{16\pi^2}8\Lambda_E
\Big[ 1-\frac{1}{s} +O(\Lambda_E/M)\Big] .
\eeq
The first derivative of
$\Sigma_H(k)$ with respect to $k_4$ is given by:
\beq\label{eq:sigma4}
\frac{\partial \Sigma_H}{\partial k_4} \Big|_{k=0} ~ = ~
-i\frac{\lambda^2}{16 \pi^2}
2 \Bigg[ \log s^2 + O(\Lambda_E/M) \Bigg] .
\eeq
We note that the Taylor expansion of the self-energy around the
on-shell point $k=0$ is not singular because the lower cut-off
$\Lambda_E/s$ acts as a regulator for the soft divergences.

First derivatives of $\Sigma_H(k)$ with respect to spatial momentum
components $k_i$ vanish due to parity.
Therefore, linear operators in the
spatial derivatives of the form $\phi^{\dagger}\partial_i\phi$
are not induced in the effective hamiltonian by loop effects
(this was implicitly assumed in eq.~(\ref{eq:longs}),
where we have done a Taylor
expansion of $\Sigma_H(k)$
to disentangle the leading terms in $H_{eff}$).

\noindent
The non-vanishing second derivatives of $\Sigma_H$ are given by:
\beq
\frac{\partial^2 \Sigma_H}{\partial k_4^2} \Big|_{k=0},~~~
\frac{\partial^2 \Sigma_H}{\partial k_i^2} \Big|_{k=0} ,
\eeq
and turn out to be proportional to $s/\Lambda_E$.
They are related respectively to the irrelevant operators
$\phi^{\dagger}\partial_4^2\phi$ and
$\phi^{\dagger}\partial_i^2\phi$.

The self-energy of the light particle
$\Sigma_L(l)$ involves a virtual
heavy pair and can be neglected
at the lowest order in $1/M$ \cite{dec}:
\beq\label{eq:sigmali}
\Sigma_L(l)~=~0 .
\eeq
The effects of heavy pairs, together with all the other subleading
effects \cite{mms}, will be treated in a subsequent work.

\noindent
The vertex correction is given by:
\beqn\label{eq:svert}
V_{\phi\phi}^a(k,k')&=&\lambda^3\int_{D3}
\frac{d^4l}{(2\pi)^4}~\frac{1}{l^2+m^2}\times
\\ \nonumber
&\times&\frac{1}{k_4+l_4-i(k+l)^2/2M}
{}~\frac{1}{k_4'+l_4-i(k'+l)^2/2M} ,
\eeqn
where $k$ and $k'$ denote respectively the momenta of the incoming
and outcoming heavy particle.
At zero external momenta:
\beq\label{eq:verte}
V_{\phi\phi}^a(0,0)~=~-\frac{\lambda^3}{16\pi^2}\Bigg[2\log s^2
+O(\Lambda_E/M) \Bigg] .
\eeq

\subsection{Fixed points}
\label{fixpo}

Since the first of eqs. (\ref{eq:flow})
for the coupling constant $\lambda$ is homogenous,
there exists a gaussian fixed point:
\beqn \nonumber
\lambda^*&=&0
\\ \nonumber
\delta M^*&=&0
\\
{m^2}^*&=&0 .
\eeqn
This fixed point is trivial, and we seek a fixed point near this one
(i.e. with $\lambda^*\ll 1$),
corresponding to a weakly interacting theory.

\noindent
In one-loop approximation,
the new coupling constant $\lambda'$ is a function of
the old coupling constant $\lambda$ only, while the new mass
parameters $\delta M'$ and ${m^2}^{'}$
depend on the original mass parameters
$\delta M$ and $m^2$ and on $\lambda$:
\beqn \nonumber
\lambda'&=&f(\lambda)
\\ \nonumber
\delta M'&=&g(\delta M,\lambda)
\\
{m^2}^{'}&=&h(m^2,\lambda) ,
\label{eq:param}\eeqn
where $f,~g$ and $h$ are given functions,
which we will compute at one loop
level later in this section.
We note that the equations of Wilson's $RG$ transformation have
the same structure as that in eqs.~(\ref{eq:param}).

\noindent
The problem is therefore
that of finding a fixed point for the coupling
constant equation:
\beq
\lambda^*~=~f(\lambda^*) .
\eeq
Once this has been found, we tune
the original mass parameters in such a way that they are unchanged
by the transformation; we have simply to solve the equations:
\beqn\nonumber
\delta M^*&=&g(\delta M^*,\lambda^*)
\\
{m^2}^*&=&h({m^2}^*,\lambda^*) .
\eeqn
Unlike the free case,
we have $\delta M^*\neq 0$ and ${m^2}^*\neq 0$, in
order to cancel the contribution
to mass renormalization coming from
integrated (high energy) modes.

Substituting eqs.~(\ref{eq:sigma4}),
(\ref{eq:sigmali}) and (\ref{eq:verte})
in the coupling constant equation (\ref{eq:flow}),
we see that the $\log s^2$ terms cancel, giving:
\beq
\lambda'~=~\lambda ,
\eeq
plus corrections of order $\Lambda_E/M$.
The coupling constant therefore does not
change with the $RG$ transformation.
This occurs because heavy pairs are
absent and the model reduces  to a free scalar
field interacting with fixed sources.
We have therefore:
\beqn\nonumber
\lambda^*&=&\lambda_0
\\ \nonumber
{m^2}^*&=&0
\\
\delta M^*&=&\frac{{\lambda^*}^2}{16\pi^2}8\Lambda_E .
\label{eq:pertfp}\eeqn
There exists a fixed point hamiltonian $H^*$ with a coupling
equal to the original one.
Therefore, if we start with an initial coupling $\lambda_0\ll 1$
at a scale $\sim M$, we remain in a region of small coupling, where
our perturbative computation is trustworthy.
We see that the formalism developed has a complete correspondence
with physical intuition. We can consider the result (\ref{eq:pertfp})
as a rigorous proof
of the existence of the static limit for our model.

\section{Matching}
\label{match}

We consider now the construction of the effective hamiltonian as
specified by eq.~(\ref{eq:minep}),
i.e. without the rescaling of the momenta and
of the fields. This is indeed the
formulation which can be more easily related to the usual approach
to the effective hamiltonians and to matching theory.
We integrate now the degrees of freedom with momenta between
$\Lambda$ and $\Lambda_E$, instead of between $\Lambda_E$ and
$\Lambda_E/s$ as we have done in the previous section.

\noindent
At leading level, the effective hamiltonian is given by:
\beqn \nonumber
&&H_{eff}\cong\int_0^{\Lambda_E} \frac{d^D k}{(2\pi)^D}
\phi^{\dagger}(k)
\Bigg[
\Bigg(1+
i\left(
\frac{\partial\tilde{\Sigma}_H}{\partial k_4}\right)_0\Bigg)ik_4+
\Big(\delta M-\tilde{\Sigma}_H(0)\Big)\Bigg]\phi(k)
\\
&+&\int_0^{\Lambda_E}
\frac{d^Dl}{(2\pi)^D}
a(-l)\Bigg[
\Bigg(1-\left(\frac{\partial\tilde{\Sigma}_L}{\partial l^2}
\right)_0\Bigg)
l^2+\Big(m^2-\tilde{\Sigma}_L(0)\Big)\Bigg]a(l)+
\label{firsteff}\\ \nonumber
&+&\int_0^{\Lambda_E}
\frac{d^Dk_1}{(2\pi)^D}...
\frac{d^Dk_3}{(2\pi)^D}(2\pi)^D\delta(k_1-k_2-k_3)
\Bigg[\lambda-\tilde{V}_{\phi\phi }^{a}(0,0,0)\Bigg]
\phi^{\dagger}(k_1)\phi(k_2)a(k_3) ,
\eeqn
where a tilde indicates the new range of integration.

\noindent
The functions
$\tilde{\Sigma}_H(0)$,
$\tilde{\Sigma}_L(0)$,
$(\partial\tilde{\Sigma}_H/\partial k_4)_0$,
$(\partial\tilde{\Sigma}_L/\partial l^2)_0$
and $\tilde{V}_{\phi\phi}^a(0,0,0)$ contain the leading effects
of the integrated modes. We present a one-loop calculation of
these functions in Sect.~\ref{pertev2}. Sect.~\ref{general}
contains a discussion and a generalization of the one-loop
results.
The functions entering in eq.~(\ref{firsteff})
can be related to the
perturbative matching constants between
the full and the static theory.
This will be done in Sect.~\ref{connec}.

\subsection{Perturbative computation}
\label{pertev2}

The Feynman rules are the same as those in Sect.~\ref{pertev}
with the only difference that now loops
are integrated in a region where
all the internal lines have momenta
between $\Lambda$ and $\Lambda_E$,
instead of between $\Lambda_E$ and $\Lambda_E/s$.

\noindent
The self-energy of the heavy scalar is given by:
\beq
\tilde{\Sigma}_H(k)=-i\lambda^2\int\frac{d^4l}{(2\pi)^4}
\frac{1}{l_4-il^2/2M}~\frac{1}{(k-l)^2+m^2} .
\label{eq:smine2}\eeq

\noindent
Let us make some qualitative remarks about the physical
meaning of this diagram (and of similar ones).
Near the upper limit of integration, the loop momentum is very large,
$l\sim\Lambda$, the shift
(\ref{eq:shift}) is irrelevant and the integrand
is similar to the corresponding one in the full theory.
On the contrary, near the lower limit of integration,
the loop momentum is very small, $l\sim\Lambda_E$. We have that
$l^2/2M\ll l_4$, and the integrand
resembles that of the static theory.
The integrand in eq.~(\ref{eq:smine2}) therefore
interpolates between the
region in momentum space
in which the heavy particle $H$ is dynamical and
the region in which $H$ is essentially static.
The amplitude therefore
includes the effects of the fluctuations
with momenta both greater and smaller
than $M$.
The transformation (\ref{eq:minep})
indeed lowers the cut-off in such a way
that we cross a physical threshold, the heavy particle mass.

\noindent
The self-energy at zero external momentum is given by:
\beq
\tilde{\Sigma}_H(0)  =  \frac{\lambda^2}{16 \pi^2}
  2M \Bigg[ \log\frac{\Lambda^2}{M^2}+1-4\frac{\Lambda_E}{M}
  +O((\Lambda_E/M)^2) \Bigg] .
\eeq
The mass renormalization of the heavy scalar $\delta M$
is related to the self-energy by:
\beq
\delta M~=~-\tilde{\Sigma}_H(0) .
\eeq
The first derivative of $\tilde{\Sigma}_H$
with respect to $k_4$ is given by:
\beq\label{eq:sigma42}
\frac{\partial\tilde{\Sigma}_H}{\partial k_4} \Big|_{k=0} ~ = ~
-i\frac{\lambda^2}{16 \pi^2}
2 \Bigg[ \log\frac{M^2}{\Lambda_E^2}-1+O(\Lambda_E/M) \Bigg] .
\eeq
In a renormalizable theory (instead of superrenormalizable),
we would have an additional term of the form $\log \Lambda^2/M^2$
at the right hand side of eq.~(\ref{eq:sigma42}).
In our model, a term of this kind appears only
in the mass renormalization, because field and coupling constant
corrections are $u.v.$ finite.
The physical origin of the two kind of logs will be discussed
in sect.\ref{general}.

One of the effects of the integrated modes is that of modifying the
normalization of the kinetic operator
given in the original hamiltonian.
This effect is quantified by
the $Z$-factor relating
the on-shell renormalized field $\phi_{OS}$ to the bare field $\phi_B$,
defined by
\beq\label{eq:defZ}
\phi_{OS}~=~\frac{\phi_B}{\sqrt{Z}} ,
\eeq
and given by:
\beq\label{eq:exprZ}
Z~=~1-i\frac{\partial \tilde{\Sigma}_H}{\partial k_4} \Big|_{k=0} .
\eeq

\noindent
The vertex correction is given by:
\beqn\label{eq:svert2}
\tilde{V}(k,k')&=&\lambda^3\int
\frac{d^4l}{(2\pi)^4}~\frac{1}{l^2+m^2}\times
\\ \nonumber
&\times&\frac{1}{k_4+l_4-i(k+l)^2/2M}
{}~\frac{1}{k_4'+l_4-i(k'+l)^2/2M} .
\eeqn
At zero external momenta:
\beq
\tilde{V}(0,0)~=~-\frac{\lambda^3}{16\pi^2}
\Bigg[2\log\frac{M^2}{\Lambda_E^2}
          +O(\Lambda_E/M)\Bigg] .
\label{eq:vert0}\eeq
In a renormalizable
theory we would have an additional term of the form
$\log\Lambda^2/M^2$ at the right hand side of eq.~(\ref{eq:vert0}),
related to the ultraviolet divergence of the
vertex correction.

The bare charge is therefore renormalized as follows:
\beq\label{eq:rencoup}
\lambda~\rightarrow~\lambda\Bigg[ 1+
\frac{\lambda^2}{16\pi^2} 2\log\frac{M^2}{\Lambda_E^2}~+O(\Lambda_E/M)\Bigg]Z
{}~=~\lambda\Bigg[ 1+
\frac{\lambda^2}{16\pi^2} 2+O(\Lambda_E/M)\Bigg] .
\eeq
The cancellation of the
$\log M^2/\Lambda_E^2$ terms is related to the
conservation of the current
$\phi^{\dagger}\phi$ in the effective theory
and is not specific of our model.

In $QED$, we would have an additional term of the form
$\beta_0\log\Lambda^2/M^2$
at the right hand side of eq.~(\ref{eq:rencoup}).
This implies that the (bare) coupling constant
of the original theory must be evolved from $\Lambda$ to $M$ only,
because it remains unchanged
as we enter the effective theory region
(i.e. the cut-off becomes smaller than $M$).

\subsection{Qualitative considerations}
\label{general}

Quite generally,
we can identify three regions of momenta which must
be integrated for the derivation of the effective hamiltonian.

1) The first one involves momenta $l$ between
the original cut-off $\Lambda$ down to a scale of the order
of the heavy particle mass:
\beq\label{reg1}
M^2~\ll~l^2~\ll~\Lambda^2 .
\eeq
In this region, the integrand
is similar to that one in the full theory,
and the mass $M$ of the heavy particle can be neglected.
We are basically computing the quantum fluctuations for a
relativistic massless particle.
The contribution to mass renormalization coming from this region,
for example, is given by:
\beq\label{eq:noscale}
\frac{\delta M}{M}~\sim~
\int\frac{d^4 l}{(l^2)^2}~\sim~\log\frac{\Lambda^2}{M^2},
\eeq
because the mass of the heavy particle acts as an infrared regulator.
It is therefore this region which produces
the $\log \Lambda^2/M^2$ term in the
mass renormalization constant of our model,
and which would give an analogous term
in $Z$ and $V$ for a renormalizable theory.
The physical origin of the logarithm is the scale invariance of a
massless field theory:
there is not any scale entering the integrand of eq.~(\ref{eq:noscale}),
so that every order of magnitude range of energies
gives the same contribution to $\delta M/M$.
These kind of logs can be
resummed with the usual technique of performing many small
cut-off lowerings, from $\Lambda$ to $\Lambda-\delta\Lambda$,
  from $\Lambda-\delta\Lambda$ to $\Lambda -2\delta\Lambda$, etc.
Region (\ref{reg1}) is, let us say, a `scaling region',
because we do not cross
any physical threshold and we are basically in a massless theory.
This region is unbounded from above and is
the one relevant to study the continuum limit of the theory.

2) The second region is the border between the full and the effective
theory, and involves momenta of the order of the mass,
\beq
l^2\sim M^2 .
\eeq
This is not a `scaling' region because we are crossing
a physical threshold, the heavy particle mass.
This region does not produce large logs because it does not extend
for many orders of magnitude
in the energy scale, but gives rise to finite
terms in the matching constants.
Crossing region 2), physics goes from a relativistic to a non
relativistic one:
there is no more enough energy to produce heavy pairs
and to give a relativistic momenta to a heavy particle.

3) The third region extends from a scale of the order of the
heavy particle mass $M$,
down to the ultraviolet cut-off $\Lambda_E$
of the effective theory:
\beq
\Lambda_E^2~\ll~l^2~\ll~M^2 .
\eeq
This region is unbounded from below
because $\Lambda_E$ may be arbitrarily small with respect to $M$.
This is again a `scaling' region, because the heavy particle
mass disappears from the dynamics. We are in a different kind of
`massless' theory with respect to region 1).
The contribution of this region to field
and vertex renormalization is
given by integrals of the form:
\beq
\int \frac{d^4l}{l_4^2~l^2}~\sim~\log \frac{M^2}{\Lambda_E^2} .
\eeq
In general, it is this region
which produces the logarithms of the form
$\log (M^2/\Lambda_E^2)$.
The origin of these logs is the scale invariance of a theory
describing massless particles interacting with fixed
(i.e. infinite mass) sources.
There is not any scale entering in the integrand, and the integral
is regulated by the heavy particle mass
and the effective theory cut-off.
These logarithms can be resummed with the usual technique, composing
small lowerings of the cut-off of the effective theory.
This region has been extensively studied in the previous section.

Up to now we have considered
a model with a heavy particle and a massless
one, but it is not difficult to extend the above considerations
to a model with, for example, two species of heavy particles
and a massless particle.
Let us assume that the
masses of the heavy particles are very different,
i.e. $M'\ll M$
(this case presents in heavy flavor physics, when both
the beauty and the charm quarks are treated in the effective theory
and their ratio $M_b/M_c$ is considered large).
In this case, there are two others relevant regions to integrate away
to derive the effective theory.
The first one lies between 1) and 2), and consists of momenta
between $M$ and $M'$,
\beq
{M'}^2\ll l^2\ll M^2.
\eeq
In this region the heavier particle behaves as a
static particle, while the lighter one behaves as a massless particle.
This is a `scaling region'.
The contributions to the renormalization constants from
this region are therefore of the form:
\beq\label{eq:hybr}
\int \frac{d^4l}{l_4^2~l^2}~\sim~\log \frac{M^2}{{M'}^2},
\eeq
because the larger mass $M$ acts as an ultraviolet
cut-off to the integral
while the smaller mass $M'$ acts as an infrared cut-off.
A log like that in eq.~(\ref{eq:hybr}) is usually called an hybrid log
\cite{eichil}.

The second region is the border between the full theory for the
lighter particle and the effective one. It consists of momenta
$l$ of the order:
\beq
l^2~\sim~{M'}^2 .
\eeq

We can extend this analysis to a theory with an arbitrary
number of heavy particles, with masses $...M''\ll M'\ll M$.
In every interval between two masses, we have a scaling region
which gives rise to a log of the mass ratio, and a boundary
region.

\subsection{Connection with usual matching theory}
\label{connec}

Let us discuss now the connection between our formulation of
the effective hamiltonians and the usual matching theory.

\noindent
Consistency requires that the amplitudes of the static theory coincide
with those of the original (full) theory at the lowest order in $1/M$.
To this end, renormalization constants for the effective theory
operators are introduced, which are determined by a set of matching
conditions. Let us consider a specific example, matching of the
heavy scalar propagator.
The bare propagators
of the full and the static theory are given, near the mass-shell,
respectively by:
\beqn\label{eq:fulstat}
i\Delta_{F}(k)&=&
\frac{-i}{k_4[1+i(\partial\Sigma_F/\partial k_4)_0]+O(1/M)}
\\ \nonumber
i\Delta_{S}(k)&=&
\frac{-i}{k_4[1+i(\partial\Sigma_S/\partial k_4)_0]-i\epsilon} ,
\eeqn
where:
\beqn\label{eq:fulstat2}
\left(\frac{\partial\Sigma_F}{\partial k_4}\right)_0
&=&- 2i\lambda^2
\int_0^{\Lambda}\frac{d^4l}{(2\pi)^4}
\frac{1}{(l^2)^2}\frac{l_4}{l_4-il^2/2M}
\\  \nonumber
\left(\frac{\partial\Sigma_S}{\partial k_4}\right)_0
&=&- 2i\lambda^2
\int_0^{\Lambda_E}\frac{d^4l}{(2\pi)^4}\frac{1}{(l^2)^2} .
\eeqn
To avoid cumbersome expressions, mass terms have been omitted in
eqs.~(\ref{eq:fulstat}).

\noindent
The matching constant $Z_m$
is defined requiring that the static propagator
coincides with the full-one at the lowest order in $1/M$:
\beq\label{eq:defmat}
i\Delta_F(k)~=~Z_m~i\Delta_S(k) .
\eeq
Using eqs.~(\ref{eq:fulstat}) and eq.~(\ref{eq:defmat}),
we derive the following expression for the matching constant:
\beq\label{eq:exprZm}
Z_m~=~1-i\Bigg[\left(
\frac{\partial\Sigma_F}{\partial k_4}\right)_0
 -\left(\frac{\partial\Sigma_S}{\partial k_4}\right)_0\Bigg] .
\eeq

\noindent
Let us consider now the renormalization of the heavy particle field
induced by our transformation. After the functional integration
has been done, the bare propagator of the heavy scalar
is given, near the mass-shell, by:
\beq
i\Delta(k)~=~\frac{-i Z}{ k_4-i\epsilon+O(1/M) } .
\eeq
where $Z$ has been defined in eq.~(\ref{eq:defZ}).

\noindent
The renormalization constant $Z$ takes into account the modification
of the field normalization as we scale down the cut-off from $\Lambda$
to $\Lambda_E$. This factor therefore includes the effects of the
fluctuations with momenta between $\Lambda$ and $\Lambda_E$ in
the normalization of the field, and must be identified with the
matching constant $Z_m$:
\beq
Z~=~Z_m .
\eeq
Using eqs.~(\ref{eq:sigma42}), (\ref{eq:exprZ}), (\ref{eq:fulstat2})
and (\ref{eq:exprZm}),
we see that the two matching constants do not coincide
completely, and the difference is given by:
\beqn\label{eq:last}
Z_m-Z&=&-i\Bigg[
\left(\frac{\partial\Sigma_F}{\partial k_4}\right)_0
              -\left(\frac{\partial\Sigma_S}{\partial k_4}\right)_0
-\left(\frac{\partial\tilde{\Sigma}}{\partial k_4}\right)_0\Bigg]
\\ \nonumber
&=&-2\lambda^2\int_0^{\Lambda_E}\frac{d^4k}{(2\pi)^4}
\frac{1}{(l^2)^2}\Bigg[ \frac{l_4}{l_4-il^2/2M}-1\Bigg] .
\eeqn
Expanding in powers of $1/M$ the heavy scalar propagator
in eq.~(\ref{eq:last}), we see that
$Z_m-Z$ contains terms of the form
$(\Lambda_E/M)^n\ll 1$ with $n\ge 1$.
$Z_M$ and $Z$ therefore coincide at the lowest order in $1/M$.
We believe that this is true at every order in
perturbation theory.
The same result holds for all the other Green functions, i.e.
differences in the matching constants arise only at order $1/M$.
The static theory is constructed dropping $1/M$ operators as well
as corrections of order $1/M$
to the matching constants of the static
operators. We conclude therefore that the transformation
(\ref{eq:minep}) reproduces
correctly the matching constants of the static theory.

\section{Conclusions}
\label{concl}

We have formulated the effective theory for heavy particles with
renormalization group techniques, by integrating away the states with
high velocity and with high virtuality.
The fixed point hamiltonian
of our transformation coincides with the static hamiltonian, and
irrelevant operators are identified with $1/M$ corrections.
Matching conditions between the full and the static theory have not
to be imposed by hand as in the usual approach, but come out
quite naturally as dynamical effects of the $RG$ flow.

Our interest was to relate,
  from first principles, quantum field theory,
a system with an infinite number of degrees of freedom, with
quantum mechanics.

The formulation of the effective theories for
heavy particles presented is non-perturbative, just like Wilson's
transformation. The prescription
for the degrees of freedom to integrate away and the equations
for the $RG$ flow indeed do not rely on any perturbative expansion.
Our transformation
can accomodate, for example, instanton effects \cite{inst},
or can be computed exactly with numerical methods.
Whenever we apply perturbation theory to it, we are able
to reproduce all the know results for the matching constants.

While this approach does not provide us
with new computational tools,
we think that it clarifies the physical content
of the effective theories for heavy particles, and constitutes
a solid framework for (future) non-perturbative computations.

\section*{Acknowledgments}

We wish to express particular thanks to Prof. G. Martinelli for
many discussions and suggestions. We acknowledge also discussions
with Profs. G. Jona, G. Parisi and M.Testa.

\newpage
\section*{Figure Captions}

\begin{enumerate}
\item[]
Fig.1: Momentum space of the heavy particle. The states with momenta
inside the white region are
integrated away, while  the states with momenta
inside the black sphere are not integrated and constitute
the dynamical variables of the effective action for heavy particles.
The black sphere is centered around the momentum of a particle at rest,
$r=(M,\vec{0})$, and has a radius $\Lambda_E\ll M$.
\item[]
Fig.2: Momentum space of the light particle. The  states with momenta
inside the white region are integrated away, while those one inside
the black sphere are not integrated and constitute the dynamical
variables of the effective action for light particles. The black
sphere is centered around the null momentum and has a radius
$\Lambda_E\ll \Lambda$.
\item[]
Fig.3: One-loop self-energy diagram of the heavy scalar.
The continuous line
represents the heavy particle, while the dotted line the light scalar.
\item[]
Fig.4: One-loop correction of the heavy-heavy-light vertex.
\end{enumerate}
\end{document}